# High-Flux Entangled Photon Generation via Clinical Megavoltage Radiotherapy Beams for Quantum Imaging and Theranostics

Gustavo Olivera[1], PhD, Bashkim Ziberi[1,2], PhD, Stephen Avery [1,3], PhD, Heth Devin Skinner[4], MD, PhD, Erno Sajo[5], PhD, Hugo Ribeiro[5], PhD, M. Saiful Huq[1,4], PhD.

[1]Quantum Theranostics, Fort Meyers, FL, USA
[2]University of Tetova, Ilinden Street nn, 1200, Tetova, R. of North Macedonia
[3]Hospital of the University of Pennsylvania, Department of Radiation Oncology, Perelman Center for Advanced Medicine, Philadelphia, PA 19104, USA
[4]Department of Radiation Oncology, University of Pittsburgh School of Medicine and UPMC Hillman Cancer Center, Pittsburgh, PA 15232, USA
[5]Department of Physics & Applied Physics, University of Massachusetts Lowell, Lowell, MA

**Corresponding author**: M. Saiful Huq (huqs@upmc.edu), Department of Radiation Oncology, University of Pittsburgh School of Medicine and UPMC Hillman Cancer Center, Pittsburgh, PA 15232, USA

**ORCHID ID**: 0000-0002-3990-2613

## Abstract

**Purpose/Objective(s):** Traditional quantum-entangled photon sources, such as spontaneous parametric down-conversion, suffer from low flux, limiting clinical translation. We investigated whether clinical megavoltage (MV) radiotherapy beams can act as dual-purpose sources, delivering therapeutic dose while simultaneously generating high-flux entangled photon pairs for quantum ghost imaging, quantum Theranostics (QTX), and related applications.

**Materials/Methods:** Monte Carlo simulations (GEANT4) modeled water-equivalent phantoms with spherical tumors (2 cm) loaded with gold nanoparticles (AuNPs, 10 mg/mL). MV beams (6, 10, 15 MV) were simulated using clinical spectra. Metrics included 511 keV photon-pair yield, positronium lifetimes (time-resolved), Doppler and energy broadening (energy-resolved), signal-to-noise ratios (SNR), and entanglement retention with depth (5–15 cm).

**Results:** Pair yields scaled with beam energy and AuNP concentration, reaching ~$10^8$ pairs per Gy per cm³ in 10 mg/mL AuNP-loaded tumors. Positronium lifetimes shift ~100 ps is associated to changes in oxygenation and reactive oxygen species, providing a biomarker. Doppler broadening (1–3 keV) and AuNP-induced line broadening (0.5–1 keV) produced spectroscopic fingerprints of electron density, tissue composition, and nanoparticle uptake. Lifetime-only QTX revealed oxygen/ROS status, Doppler-only QTX revealed composition/density, and the combined approach enabled a full Theranostics map. Depth-dependent simulations showed voxel SNR ~$10^2$. This is a conservative estimate accounting for photon escape, coincidence efficiency, and entanglement gating.

**Conclusion:** Clinical MV radiotherapy beams are a practical high-flux entangled photon source. By integrating time-resolved positronium lifetimes with energy-resolved Doppler and Compton spectroscopy, QTX enables simultaneous functional mapping of tumor microenvironment and spectroscopic determination of tissue composition during therapy. This supports nanoparticle-optimized phantom development and establishes a high-energy entanglement platform, extending quantum technologies from optical (eV) to therapeutic (hundreds of keV) regimes. Such a platform could unlock high-yield entanglement, super-resolution ghost imaging, advanced quantum sensing, and probes of fundamental physics, while offering transformative opportunities in precision medicine.

# Introduction

Advances in radiotherapy and multimodal imaging have markedly improved cancer treatment, yet fundamental limitations remain in our ability to dynamically monitor tumor response and microenvironmental changes in real time during therapy. Current imaging modalities, including positron emission tomography (PET), computed tomography (CT), and magnetic resonance imaging (MRI), provide complementary anatomical and functional insights but are constrained by poor temporal resolution, indirect biochemical specificity, and their inability to seamlessly integrate into the radiotherapy workflow [1]. As a result, adaptive treatment guidance remains reliant on static or delayed imaging data, limiting the precision and personalization of therapy.

Quantum technologies have emerged as a disruptive frontier for biomedical imaging. In particular, entangled photon pairs offer non-classical correlations in energy, time, and momentum that can be exploited for quantum-enhanced imaging, spectroscopy, and sensing [2-5]. Techniques such as ghost imaging and entangled photon spectroscopy have demonstrated the ability to surpass classical resolution and sensitivity limits. However, current laboratory-scale sources of entangled photons are predominantly based on spontaneous parametric down-conversion (SPDC). SPDC generates extremely low number of entangled pairs (1 pair per $10^{10}$ photons) [6, 7], much lower than QTX generates (1 pair per 100 photons). The photon energy range of SPDC sources (typically less than 3 eV, optical and near-infrared) also limit tissue penetration and relevance for high-energy oncology applications.

Clinical megavolt (MV) radiotherapy beams represent a unique and previously unexplored potential solution to these bottlenecks. When interacting with tissue, MV photons produce secondary electron–positron pairs that subsequently annihilate into 511 keV photon pairs. These pairs exhibit angular, temporal, and polarization correlations consistent with quantum entanglement, as recently demonstrated in PET imaging [8]. The use of MV radiotherapy beams could unlock a source of entangled photons at fluxes several orders of magnitude greater than laboratory entanglement sources, and it is inherently integrated into the therapeutic dose delivery process. The use of high-Z nanoparticles, such as gold (AuNPs), could further amplify pair production cross-sections, and enhance both photon yield and entanglement fidelity [9-11].

This work introduces the concept of Quantum Theranostics (QTX), which leverages entangled photon pairs generated during radiotherapy for simultaneous treatment and quantum-enhanced imaging. We present Monte Carlo simulations (Geant4) [12] that model the generation, propagation, and detection of entangled 511 keV photon pairs in tissue-equivalent phantoms with and without AuNPs. Primary metrics are photon-pair yield, angular correlation functions, and entanglement coherence as a function of beam energy and depth. Positronium lifetime distributions are treated as a secondary biomarker, complemented by energy-resolved Doppler and Compton spectroscopy, which together provide a complete Theranostics framework. By bridging precision medicine, quantum science, nanotechnology, AI, quantum computing, QTX approach could establish a new class of integrated Theranostics technologies that advance quantum medicine such as quantum-enabled biomedical imaging and beyond. We emphasize that photon yield estimates and voxel-level SNR values are on the order of $10^8$ pairs per Gy per cm$^3$ and $10^2$ per voxel, respectively, and represent conservative simulation-based values anchored to prior experimental measurements.

# 2. Methods

## 2.1 Monte Carlo Phantom Model

Monte Carlo simulations were performed in Geant4 v11.2 [12, 13] using the G4EmStandardPhysics_option4 list, which accurately models electromagnetic interactions at MV energies (pair production, Compton scattering, photoelectric effect). The geometry consisted of a $30 \times 30 \times 30$ cm³ water-equivalent cube containing a 2-cm-diameter spherical tumor centered in the volume. Tumor voxels were loaded with gold nanoparticles (AuNPs; 10 mg/mL equivalent), represented as a homogeneously distributed atomic mixture. This configuration emulates a high-Z–enriched microenvironment to enhance pair production; delivery mechanisms (e.g., direct injection or systemic uptake) were not modeled.

The phantom was voxelized into 0.5 mm³ cubic voxels, and tallies included energy deposition, photon counts, spatial coordinates, and polarization states. Particle transport used 1 keV production/transport thresholds for photons and electrons.

While AuNPs are referred to in the text, they were modeled here as a homogeneous atomic gold distribution within the tumor voxel, rather than discrete nanoparticles of specific size. In this representation, particle size does not influence the simulation outcome, and the reported dose deposition may overestimate the realistic case where nanoparticles exist in discrete clusters. The purpose of this approximation is to establish an upper-bound estimate of pair-production yield and feasibility for entangled-photon generation.

### 2.1.1 Dose considerations

In Geant4, energy deposition can be tallied as dose-to-medium ($D_m$) or converted to dose-to-water ($D_w$). At megavolt photon energies, $D_w$ and $D_m$ differ by only a few percent in soft-tissue–equivalent media [14]. $D_w$ remains the clinical reporting standard, but $D_m$ more directly represents the underlying stopping power ratios and microscopic energy deposition processes within the simulated material. For this study, tallies were recorded in $D_m$, with the understanding that the macroscopic difference to $D_w$ is modest. However, when high-Z nanoparticles are present, local microscopic dose distributions are significantly perturbed, as photoelectric and pair-production events near AuNP surfaces deposit energy within nanometer-scale regions. These "nano-dose" effects yield dose enhancement factors (DEF) on the order of 8–10 times in the immediate nanoparticle vicinity, far greater than $D_w$ versus $D_m$ differences, and constitute a biologically significant effect relevant to QTX.

### 2.1.2 Beam Modeling

Clinical megavoltage photon beams at 6, 10, and 15 MV were modeled using clinical LINAC spectra. Field size, SAD = 100 cm, and lateral fluence profiles were matched to clinical treatment beams. Transport through the phantom produced electron–positron pairs via nuclear-field pair production; subsequent positron annihilation yielded correlated (entangled) 511 keV photon pairs within the tumor volume. For each voxel, we tallied secondary-photon trajectories, energies, and polarization states. Entanglement preservation was quantified by the distribution of opening-angle deviations from 180° (Δθ); positronium lifetime distributions were computed as a secondary, exploratory metric of microenvironment sensitivity.

### 2.1.3 Positronium states

Both para-positronium (2γ, ~125 ps) and ortho-positronium (3γ continuum, strongly quenched in tissue) were included conceptually in the modeling. Para-Ps contributes primarily to time-domain lifetime sensitivity, while ortho-Ps quenching influences both lifetime shortening and Doppler-broadening spectra, providing complementary observables in QTX.

### 2.1.4 Simulation Parameters Summary

This framework allows rigorous evaluation of MV radiotherapy as a high-flux source of entangled photons, bridging therapeutic dose delivery and quantum-enhanced imaging, and establishing the foundation for optimized nanoparticle phantoms and broader quantum applications. A summary of phantom paramteres used for Monte Carlo simulations performed in Geant4 is given in table 1.

| Parameter | Value / Range |
|---|---|
| Phantom | Water-equivalent, 30 × 30 × 30 cm³ |
| Tumor | 2 cm diameter, 10 mg/ml AuNP volume |
| Voxel size | 0.5 mm³ |
| Beam energies | 6, 10, 15 MV |
| Particle cutoffs | 1 keV for photons/electrons, 1 μm step size |
| Photon pair energy | 511 keV |
| Detector radius | 10–50 cm |
| Coincidence window | 200 ps |
| Monte Carlo events | $10^9$ primary photons per simulation |

**Table 1**: Summary of phantom parameters used for Monte carlo simulations.

## 2.2 Detectors

To assess the detectability and utility of entangled-photon signals, we implemented a virtual detector array surrounding the phantom at a 10–50 cm radial distance. The array comprised four idealized channels: TES (energy resolution 0.1–1 keV, modeled as ideal energy absorbers) [15], SNSPDs (timing resolution <150 ps with polarization sensitivity) [16], SPADs (high-rate counting, $\geq 10^8$ s$^{-1}$) [17], and Timepix sensors (spatial sampling <1 mm) [18]. Coincidences were formed within a 200 ps window consistent with high-speed SNSPD performance. Transport included ballistic transmission, absorption, single Compton, and multiple scattering. For each voxel, we computed SNR, entanglement-coherence retention (e.g., opening-angle/polarization correlations), and energy-resolved spectroscopic metrics.

## 2.3 SNR computation

Voxel-level SNR was determined through a chain of simulated factors: entangled pair yield, annihilation site distribution (inside vs. outside tumor), photon escape probability, coincidence efficiency, and entanglement retention. Reported SNR values also incorporate reductions due to coincidence gating and image reconstruction. These represent conservative, simulation-based estimates anchored to published detector performance and are intended to demonstrate the feasibility of high-fidelity image recovery at depth.

## 2.4 Detector Configuration

To capture entangled-photon signals and extract useful quantum correlations, QTX will mount hybrid detectors on compact robotic arms, with configurations prescribed by forthcoming Monte Carlo studies. Patient- and fraction-specific simulations will choose detector geometries that maximize coincidence efficiency and angular coverage while respecting gantry/couch clearances, minimizing scatter, and avoiding the therapeutic beam; intra-fraction adjustments will follow a precomputed Monte Carlo repositioning plan. Each detector type in the array plays a specialized role:

- Transition Edge Sensors (TES): measure energy with resolution of (0.1–1.0 keV) to detect subtle Doppler and Compton shifts. TES requires cryogenic operation (~100 mK), supported by dilution refrigerators or compact adiabatic demagnetization systems [15, 19].

- Superconducting Nanowire Single Photon Detectors (SNSPDs): deliver <150 ps timing precision for time-of-flight and coincidence detection, while also capturing polarization states to verify entanglement fidelity [16].

- Single Photon Avalanche Diodes (SPADs): provide high-speed photon counting ($\geq 10^8$ counts/s), useful for rapid absorption imaging and triggering gated detection [17].

- Timepix detectors: offer <1 mm spatial resolution for reconstructing ghost images and anatomical boundaries [18, 20].

Figure 1 illustrates the spatial arrangement and geometry of the detectors. These detectors will be evaluated to identify the optimal configuration for real-time acquisition of quantum observables: energy, temporal resolution, polarization, and spatial position; enabling high-fidelity imaging and in situ spectroscopic mapping.

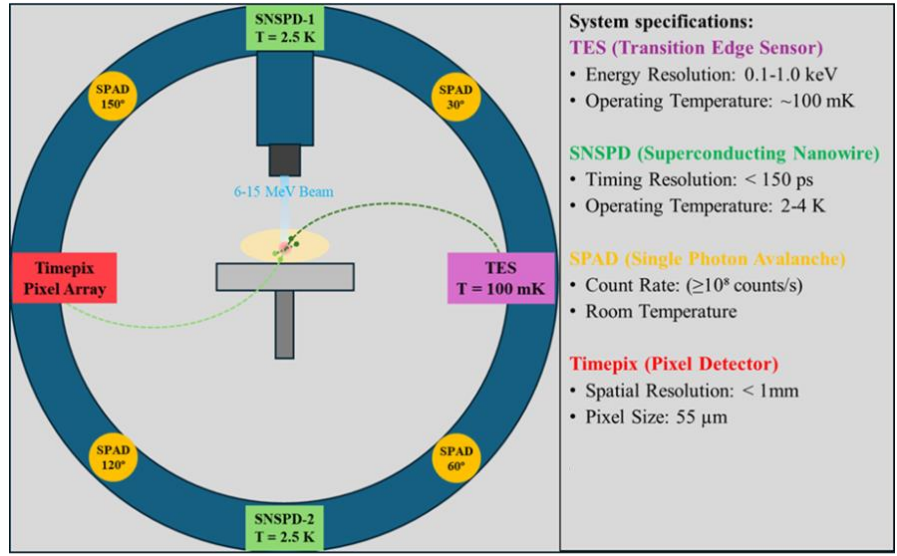

**Figure 1:** Schematic of the QTX system, illustrating the detector locations relative to the beam and patient geometry.

## 2.5 Photon Interaction Pathways for Imaging and Spectroscopy

Entangled photon generation is illustrated in Figure 2. Once produced, photon pairs propagate through the tumor and surrounding tissue, undergoing four principal interaction pathways, each contributing uniquely to the resulting imaging and spectroscopic signals. These interaction modes are summarized in Table 2.

In the proposed QTX platform, entangled photon pairs are generated during radiotherapy by introducing high-Z nanoparticles into the tumor microenvironment to act as interaction catalysts. Gold nanoparticles (AuNPs) are selected for their high atomic number (Z = 79) and established biocompatibility. A suspension concentration of 10 mg/mL, functionalized with anti-EGFR antibodies, is assumed to enable selective uptake in epithelial tumors [9, 10, 21, 22]. Intratumoral injection would also be conceivable for tumors such as breast, prostate and other solid tumors.

Pair production is initiated when high-energy photons from a clinical linear accelerator (LINAC) interact with gold atoms, predominantly through nuclear field-induced conversion. This process generates electron-positron pairs within the tumor volume. The positrons undergo rapid annihilation with local electrons, producing 511 keV photon pairs in an entangled state, which form the quantum basis for QTX imaging and therapeutic applications.

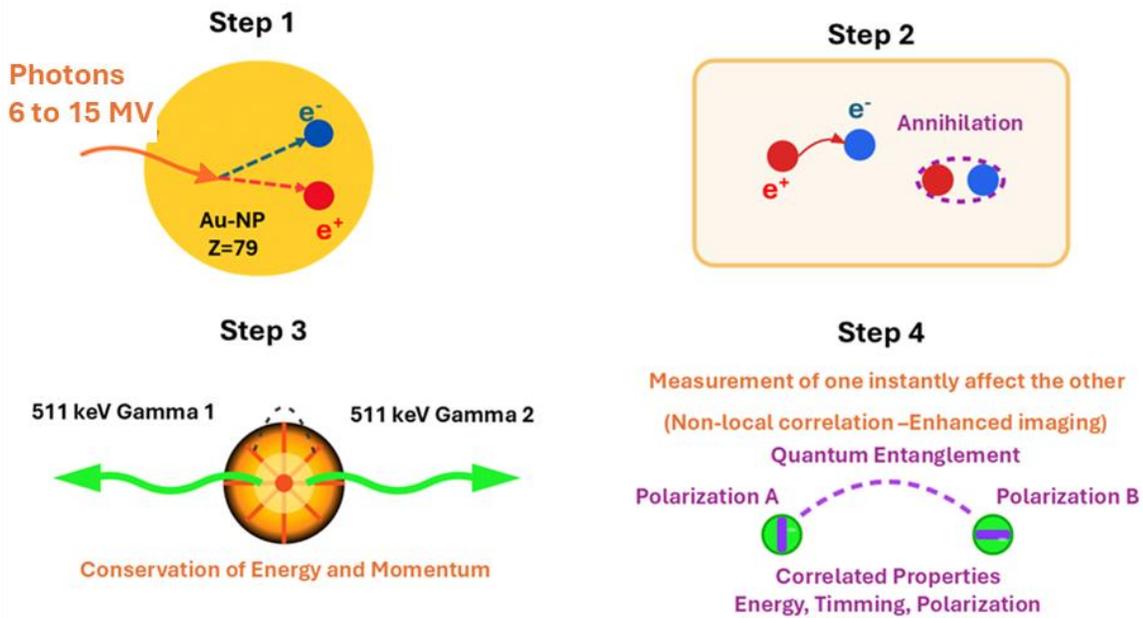

**Figure 2**: Schematic of entangled photon generation in Quantum Theranostics: (1) Pair production primarily in gold nanoparticles; (2) Annihilation in tumor tissue; (3) Annihilation photons with 511 keV; (4) Detection of QIEP with entanglement in energy, momentum, and polarization.

| Scenario | Physical Process | Imaging Contribution | Imaging Modality (Quantum/Classical) | Spectroscopic Information |
|---|---|---|---|---|
| **(1) Two Ballistic Photons** | Avoid scattering/absorption | High-resolution structural imaging, tumor boundaries (<1 mm) | Ghost Imaging [3] /Direct projection | None (511 keV unchanged) |
| **(2) Absorption Events** | Photoelectric in dense/high-Z regions | Absorption maps highlighting nanoparticle-rich zones | Ghost Imaging (idler inference)/Shadow | Energy loss indicating density |
| **(3) Compton Scattering** | Tied to electron density, enhanced in ROS (Reactive Oxygen Species) | Spectroscopic mapping of ROS/heterogeneity via energy shifts | Ghost Imaging (idler correlation)/Scatter mapping | Doppler/Compton shifts (13–171 keV) revealing ROS/heterogeneity |
| **(4) Multiple Scattering/Noise** | Dual scatter/absorption creates noise; preserved entanglement in low-scatter events | Filtered for signal clarity; correlated reconstruction if entanglement retained | Ghost Imaging (filtered)/Noise subtraction | Mixed energy for suppression; dual shifts for electron density/ROS if preserved |

**Table 2**: Photon Interaction, imaging and spectroscopy scenarios: 1: Both photons escape without interaction (ballistic, maintaining 511 keV energy, polarization, and ±150 ps timing). Scenario 2: At least one photon absorbed. Scenario 3: At least one photon undergoes Compton scattering. Scenario 4: Both photons interact via scatter or absorption, disrupting entanglement correlations.

From the Monte Carlo simulations, the following voxel-level observables were extracted:
- Photon-pair yield: scaled as a function of MV beam energy and AuNP loading.
- Angular correlations: deviations from 180°, used as a proxy for entanglement fidelity.
- Signal-to-noise ratio (SNR): evaluated as a function of tissue depth (5-15 cm).
- Entanglement-coherence retention: quantified per voxel to assess feasibility of quantum imaging.
- Positronium lifetime distributions: included as a secondary metric, with interpretive implications deferred to the Discussion section.

# 3. Results

## 3.1 Photon Pair Yield and Voxel-Level SNR

Monte Carlo simulations predicted that AuNP-loaded tumor volumes exposed to MV beams produce entangled photon-pair yields on the order of $10^8$ pairs per Gy per cm³. For a standard 2 Gy fraction, this corresponds to the order of $10^9$ photon pairs within the tumor volume. These values are consistent with published pair-production cross-sections and represent conservative simulation-based estimates rather than direct measurements (Figure 3A). It is important to note that Geant4 does not directly simulate entanglement phenomena. Instead, photon-pair yields and annihilation spectra were tallied classically, and estimates of entanglement coherence retention were applied through analytical and literature-based models. As a result, the entanglement-related observables represent modeled overlays on top of Geant4 photon transport outputs, not native Geant4 tallies.

Depth-dependent voxel-level SNR was computed from the simulated chain of processes (pair yield, annihilation site, photon escape, coincidence efficiency, entanglement retention, and reconstruction). Values were on the order of $10^2$, with approximate results of 158 at 5 cm, 119 at 10 cm, and 86 at 15 cm depth (Figure 3B). These values incorporate photon escape, coincidence efficiency, and entanglement gating, and therefore represent conservative predictions anchored to detector performance.

## 3.2 Spectroscopic Capabilities

Energy-resolved measurements with Transition Edge Sensors (TES) enable several spectroscopic analyses (Figure 3C):

- **Doppler Broadening Analysis**: Residual positron-electron momentum manifests as spectral broadening on the order of a few keV, revealing variations in local electron density and tissue biochemistry. Even small redshifts (~1–2 keV) are resolvable, providing a sensitive probe of microenvironmental heterogeneity [2, 4, 5, 23, 24].
- **Compton Scattering Signatures**: Scattering-induced energy shifts (13–171 keV) can be measured with sub-keV TES precision, enabling discrimination of photon paths and tissue composition [2, 4, 5, 23, 24].
- **Nanoparticle Interaction Effects**: AuNPs broaden the 511 keV line by 0.5–1 keV, allowing quantitative estimation of nanoparticle uptake via adjoint transport computations.

These capabilities demonstrate that QTX can simultaneously provide high-resolution structural imaging and functional spectroscopic mapping of the tumor microenvironment.

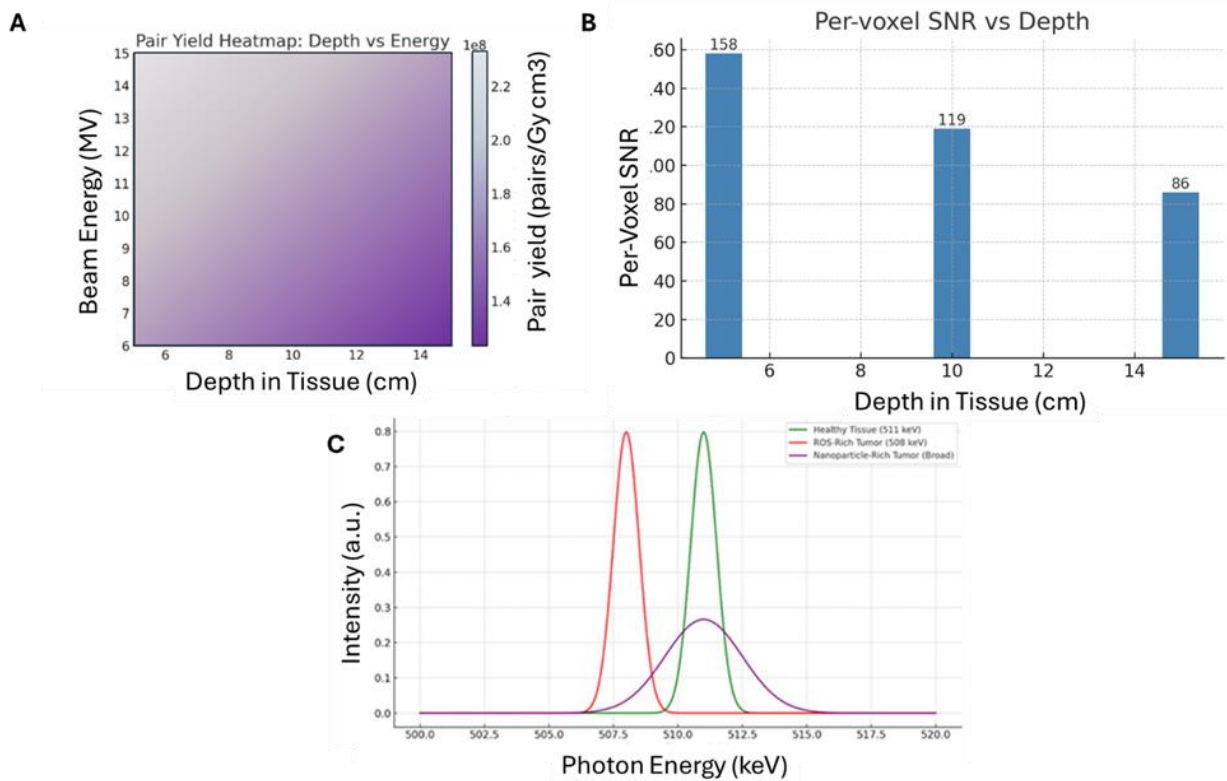

**Figure 3: A**: Heatmap of Monte Carlo–derived entangled photon pair yields for a 1 cm radius tumor loaded with 10 mg/mL AuNPs. This approximates the order of magnitudes for the yield of pairs generated. **B:** Per-voxel signal-to-noise ratio (SNR, 1 mm³ voxels) at depths of 5, 10, and 15 cm, showing values of ~158, 119, and 86 respectively. These results confirm that depth-dependent attenuation and decoherence increase, but do not eliminate, entanglement-enhanced imaging capability, with SNR consistently above the 25–100 range required for high-fidelity reconstruction. **C:** Spectral shifts in energy-resolved detection normalized intensity versus photon energy (500–520 keV) qualitatively describing different tissue types (green, 511 keV), ROS-rich tumor (red, 508 keV), and nanoparticle-rich tumor (purple, broad), highlighting energy shifts and broadening effects resolvable by TES detectors.

## 3.3 Quantum Ghost Imaging and Correlation Recovery

Ghost imaging in QTX leverages entangled annihilation photon pairs generated directly within the tumor during MV irradiation. One photon of the pair (the signal) passes through tissue, where it may undergo scattering, absorption, or Doppler shifts. Its entangled partner (the idler) exits without interaction and is recorded by a pixelated reference detector (Timepix). The signal is measured by bucket detectors (SNSPD or SPAD), which provide timing and coincidence information, and a fraction of signals are routed to TES detectors for energy-resolved spectroscopy. By correlating the idler's spatial position with the signal's detection under coincidence gating (Δt < 200 ps), the system reconstructs both structural ghost images and spectroscopic maps from the same set of annihilation pairs. Schematics of quantum Ghost Imaging is given in Figure 4.

### 3.3.1 Imaging capability

Monte Carlo simulations predict that ghost imaging can reconstruct tumor structure at sub-millimeter resolution, consistent with the 0.5–1 mm³ voxel size of the phantom. Voxel-level SNR values were as in figure 3B. These levels are above the ~25–100 range typically required for high-fidelity image recovery, supporting the feasibility of structural ghost imaging under clinical beam conditions.

### 3.3.2 Entanglement survival

Despite photon scattering and absorption, a large fraction of entanglement was preserved at clinically relevant depths. Correlation fidelity remained ~80% at 5 cm, ~70% at 10 cm, and ~60% at 15 cm. This robustness ensures that sufficient quantum correlations survive to support both ghost imaging and spectroscopic recovery.

### 3.3.3 Spectroscopic recovery

Energy-resolved TES simulations showed that the QTX platform can detect several characteristic spectral features. Doppler broadening of ~507–515 keV reflects local electron density; Compton shifts of ~340–498 keV reveal tissue composition and ROS heterogeneity; and AuNP-induced broadening of ~510–512 keV indicates nanoparticle uptake (Fig. 3C). These energy-domain features, when co-registered with ghost images, provide a functional map of the tumor microenvironment.

In summary, Together, these results demonstrate that QTX ghost imaging can achieve sub-millimeter structural resolution, preserve quantum correlations at depth, and recover spectroscopic signatures of tissue composition, ROS activity, and nanoparticle distribution. All values represent conservative, simulation-based estimates. This dual-domain capability (time-resolved positronium lifetime and energy-resolved Doppler/Compton spectra) distinguishes QTX from conventional PET or optical ghost imaging approaches.

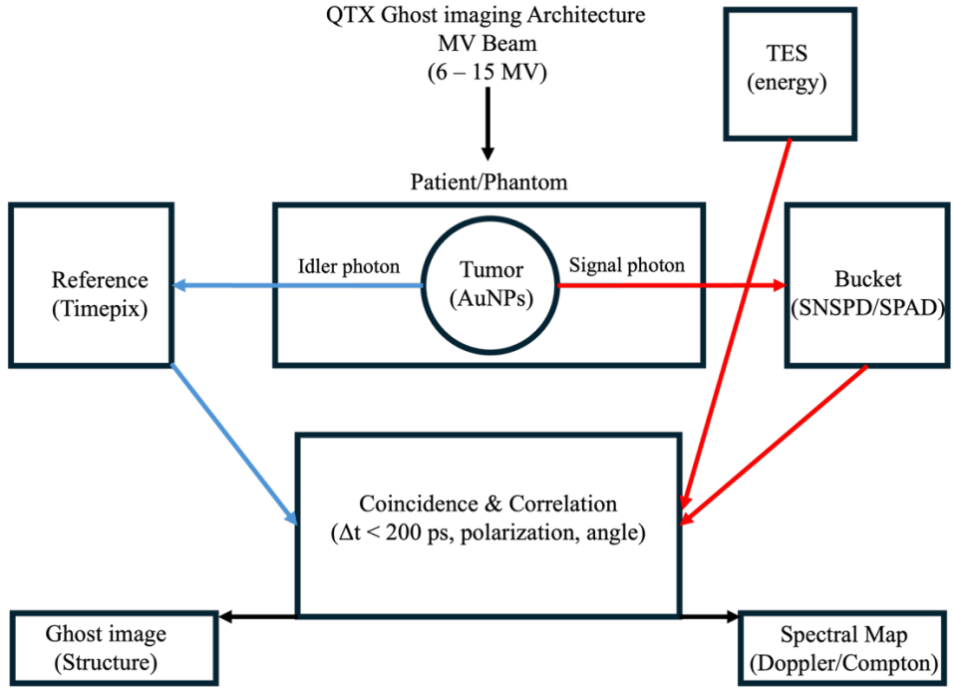

**Figure 4:** QTX ghost imaging architecture. A high-energy clinical MV beam (6–15 MV) irradiates the patient or phantom containing AuNP-loaded tumor tissue. Annihilation photon pairs are generated: the idler photon (blue) is routed to a Timepix detector for spatial reference, while the signal photon (red) is detected by a bucket arm consisting of superconducting detectors (SNSPD/SPAD) and energy-resolving TES. Coincidence and correlation analysis (Δt < 200 ps, polarization, and angular matching) reconstructs both ghost images of structural features and spectral maps of tumor microenvironment (via Doppler and Compton shifts). This dual-channel entanglement-based framework integrates structural imaging with spectroscopy for quantum theranostics.

## 3.4 Positronium Sensitivity

Positronium lifetime behavior was estimated heuristically as a probe of microenvironmental sensitivity [13, 23, 25-39]. We did not simulate positronium formation, state populations, or annihilation lifetimes in Geant4. Instead, we applied analytical mappings informed by published positronium spectroscopy to the annihilation-photon transport results in order to approximate lifetime distributions as functions of oxygenation, reactive oxygen species, electron density, and nanoparticle content. Under these assumptions, lifetime shifts on the order of 100 ps are plausible for the scenarios studied. These values are conservative design estimates, not direct simulations or measurements, and they are included to demonstrate how a time domain channel can be integrated into QTX and to guide future experimental validation.

Positronium ($e^+e^-$) formation is well studied in PET: its mean lifetime reflects microstructural voids and oxygenation, making it a molecular biomarker. However, standard PET only detects the 511 keV annihilation pair, and lifetime recovery requires a tracer-specific prompt gamma in coincidence with the annihilation photons. This severely limits event statistics. First-in-human data already show shorter lifetimes in glioblastoma than in healthy brain/salivary glands [YY], but with $^{68}$Ga only ~1.3% of decays emit a prompt gamma, making useful lifetime events >4,000 times rarer than standard γγ PET. Total-body PET plus β+γ emitters (e.g., $^{44}$Sc) can partially overcome this, providing sensitivity increases of ~28–87 times [40]. Modern scanners such as the UPenn PennPET Explorer have demonstrated the hardware capability, enabling sub-mm images and <10 ps lifetime precision.

QTX overcomes these PET bottlenecks by embedding positronium detection directly into the therapeutic irradiation process, eliminating the dependence on rare β+γ isotopes. In QTX, positronium lifetime maps are acquired in parallel with energy-resolved annihilation spectra (500–520 keV) using TES detectors. This integration allows lifetime imaging to be performed at therapeutic dose levels without additional tracers or extended scan times.

Importantly, positronium sensitivity in QTX is not an isolated channel but part of a dual-domain framework. The time domain (lifetime shortening due to oxygenation and ROS) complements the energy domain (Doppler and Compton broadening signatures of tissue composition and density). Together, these observables yield a co-registered molecular and structural map of the tumor microenvironment, a capability not achievable with PET alone.

## 3.5 Depth-Dependent Performance

Depth-dependent Monte Carlo simulations further demonstrated robust survival of entanglement, consistent with corresponding SNR estimates and supporting the feasibility of in situ quantum-enhanced imaging and spectroscopy.

To quantify voxel-level signal-to-noise, we adopted an average annihilation photon–pair yield of ~$1.8 \times 10^8$ pairs/Gy·cm³ for a 1 cm³ tumor loaded with 10 mg/mL AuNPs (Figure 3A). This baseline, derived from Geant4 simulations across 5–15 cm depths, implies that a standard 2 Gy fraction produces ~$1.3 \times 10^9$ entangled pairs within the tumor volume. Accounting for coherence retention, photon escape, and detector efficiency, the resulting per-voxel SNR falls in the 80–150 range depending on depth (Figure 3B). Importantly, these pairs are generated locally within the irradiated tumor and retain sufficient entanglement to support quantum image reconstruction [41]. These results confirm and extend those described in Section 3.1 and again represent conservative simulation-based estimates.

A defining innovation of QTX is its quantum ghost imaging framework. Here, entangled idler photons act as non-interacting proxies to reconstruct the fate of their signal partners (QIEPs) traversing tissue. Even when the signal photon undergoes scattering, absorption, or Doppler shift, the idler may preserve correlations sufficient to recover structural and spectroscopic information. While high yields suggest imaging may be feasible even without nanoparticles, this approach requires additional innovative use of current detectors and extensive experimental validation.

# 4. Discussion

## 4.1 Technical and Physical Challenges

While the Results demonstrate the feasibility of using MV radiotherapy as a high-flux source of entangled photons, several key implementation challenges must be addressed for clinical translation:

- *Cryogenic Detector Integration*: TES detectors require operation at ~100 mK. Integrating cryogenic systems safely in clinical treatment rooms poses both engineering and logistical challenges.
- *Quantum Coherence in Tissue*: Biological tissues scatter and absorb photons, potentially degrading entanglement. The survival of quantum correlations for deep-seated tumors requires further modeling and experimental validation.
- *Robotic Arm Synchronization*: Dynamic repositioning of detector arrays during treatment must be precise to avoid interference with beamlines or patient workflow.
- *Photon-Pair Discrimination & Throughput*: High photon flux generates massive data rates; custom Quantum Machine Learning (QML) pipelines[42, 43] may be required to identify entangled pairs in real time.
- *Additional Decoherence Sources*: Variations in refractive index, source imperfections, and environmental noise may further degrade entanglement beyond tissue-induced effects.

These challenges highlight the need for interdisciplinary development spanning radiotherapy physics, quantum detector engineering, and real-time signal processing. Table 3 summarizes the principal challenges for QTX, spanning technical, physical, and systems-integration domains. These implementation barriers, along with their potential clinical implications, are outlined below.

| Challenges | Description |
|---|---|
| **Cryogenic Detector Integration** | TES detectors require operation at ~100 mK; implementing cryogenics in treatment rooms poses safety and logistical hurdles. |
| **Quantum Coherence in Tissue** | Biological tissues scatter and absorb; coherence preservation (for deep seated tumors) must be validated in realistic models. |
| **Robotic Arm Synchronization** | Real-time, precise adjustment and alignment of detector arms is needed without interfering with beamlines or patient workflow. |
| **Photon-Pair Discrimination & Throughput** | Accurate identification of entangled photon pairs demands high-speed processing and custom Quantum Machine Learning (QML) filtering to handle massive data rates. |
| **Additional Decoherence Sources** | Uncertainties in refractive index, source imperfections causing polarization variations, and noise from thermal radiation or electromagnetic interference may degrade coherence beyond tissue effects. |

**Table 3**: Some of the Key Challenges for QTX Clinical Implementation.

## 4.2 Opportunities for Optimization

In parallel with the technical challenges, the QTX platform offers multiple avenues for optimization and further investigation:

- **Nanoparticle Engineering:** Beyond conventional AuNPs, alternative high-Z or hybrid designs such as perovskite-based, bismuth-containing, or core–shell architectures may enhance entangled-photon yield and coherence. Tailored surface functionalization could further promote selective tumor uptake or microenvironmental targeting.
- **Beam Profile Optimization:** Adjusting LINAC beam energies, modulation schemes, or field geometries has the potential to increase pair-production efficiency and improve uniformity of entangled-photon distribution across the tumor volume.
- **Detector Array Design:** Refinement of detector geometry, angular coverage, and coincidence timing parameters could maximize usable signal while reducing noise and decoherence.
- **Monte Carlo and Adjoint Modeling:** Systematic computational studies varying tumor size, nanoparticle concentration, and tissue composition can inform experimental design and optimize detector placement strategies.
- **Functional and Spectroscopic Imaging:** Leveraging Doppler shifts, Compton scattering signatures, and positronium lifetimes may enable real-time mapping of reactive oxygen species, tissue density heterogeneity, and nanoparticle uptake.
- **Hybrid Experimental and Theoretical Studies:** Bench-top validation of high-flux entangled-photon production using MV sources and nanoparticle phantoms will provide experimental benchmarks to test Monte Carlo predictions and guide the design of future in vivo studies.

Together, these opportunities highlight a clear translational pathway for QTX, with optimization spanning nanoparticle engineering, radiotherapy beam design, detector architectures, and hybrid experimental and computational validation frameworks.

## 4.3 Broader Implications

Beyond adaptive radiotherapy, QTX introduces a new class of integrated quantum theranostic technologies. For the first time, high-flux entangled photons are produced directly at megavolt therapeutic energies, extending quantum imaging and sensing capabilities from the optical and PET domains into the clinical radiotherapy regime. This represents a versatile platform not only for advanced ghost imaging and super-resolution reconstruction, but also for quantum sensing and fundamental physics studies at unprecedented energies.

Clinically, QTX enables simultaneous tumor treatment and real-time mapping of the tumor microenvironment. The dual-domain sensitivity includes time domain positronium lifetimes that reflect oxygenation and ROS activity, and energy domain Doppler and Compton spectra that reflect tissue composition and nanoparticle uptake. Together, these provide biochemical specificity alongside sub-millimeter structural resolution. This integration of imaging and therapy within the same irradiation session distinguishes QTX from both conventional PET and prior ghost imaging approaches.

By combining entangled-photon detection, ghost imaging, in situ spectroscopy, and quantum machine learning (QML) based reconstruction, QTX defines a framework in which every therapeutic beam interaction doubles as a measurement event. In this vision, each treatment fraction becomes not only therapeutic but also diagnostic, dynamically refining and personalizing care.

The feasibility results presented here provide early justification for targeted research and development. Immediate priorities include: (i) experimental validation of photon yields and entanglement coherence under MV beams, (ii) development of nanoparticle and tissue-mimicking phantoms for high-entanglement photon generation and detector-response testing, and (iii) advancement of image-reconstruction and spectroscopic pipelines. These steps will lay the foundation for preclinical studies and ultimately clinical translation.

# 5. Conclusion

This work establishes the conceptual and computational foundation for Quantum Theranostics (QTX), a platform that integrates entangled-photon generation with in situ imaging and spectroscopy during MV radiotherapy. Monte Carlo studies show that therapeutic beams interacting with high-Z nanoparticles can produce entangled photon pairs at clinically significant fluxes, with sufficient SNR and correlation fidelity to support ghost imaging and spectroscopic recovery at depths up to 15 cm.

- A defining feature of QTX is its dual-domain sensitivity:
  Time domain positronium lifetimes provide contrast related to oxygenation, ROS levels, and tissue microenvironment.

- Energy domain Doppler and Compton shifts reveal tissue composition, heterogeneity, and nanoparticle uptake.

Together, these complementary observables enable simultaneous structural imaging and biochemical mapping within the same therapeutic fraction.

If successfully developed, QTX would transform every treatment session into both a therapeutic and diagnostic event, offering real-time, adaptive, and personalized cancer care. Such a platform could unlock powerful applications, including ghost imaging with super-resolution, advanced quantum sensing, and novel probes of fundamental physics, while providing transformative opportunities in oncology and beyond. The feasibility results presented here justify targeted experimental validation of photon yields, entanglement survival, and detector performance, which will form the next step toward preclinical and clinical implementation.

# 6. References

1. Phelps, M.E., *PET: Molecular Imaging and Its Biological Applications*. 2004, New York, NY: Springer.
2. Al-Sheibani, Z.T., et al., *Detection of line shape parameters in normal and abnormal biological tissues.* Iraqi Journal of Physics, 2012. **10**(17): p. 77-82.
3. Lemos, G.B., et al., *Quantum imaging with undetected photons.* Nature, 2014. **512**(7515): p. 409-12.
4. Liu, G., et al., *Applications of positron annihilation to dermatology and skin cancer.* physica status solidi (c), 2007. **4**(10): p. 3912-3915.
5. Eshun, A., et al., *Entangled Photon Spectroscopy.* Acc Chem Res, 2022. **55**(7): p. 991-1003.
6. Evans, P.G., et al., *Bright source of spectrally uncorrelated polarization-entangled photons with nearly single-mode emission.* 2010, arXiv.
7. Burnham, D.C. and D.L. Weinberg, *Observation of Simultaneity in Parametric Production of Optical Photon Pairs.* Physical Review Letters, 1970. **25**(2): p. 84-87.
8. Watts, D.P., et al., *Photon quantum entanglement in the MeV regime and its application in PET imaging.* Nat Commun, 2021. **12**(1): p. 2646.
9. Ngwa, W., et al., *Targeted radiotherapy with gold nanoparticles: current status and future perspectives.* Nanomedicine, 2014. **9**(7): p. 1063-1082.
10. Butterworth, K.T., et al., *Physical basis and biological mechanisms of gold nanoparticle radiosensitization.* Nanoscale, 2012. **4**(16): p. 4830-8.
11. Brivio, D., E. Sajo, and P. Zygmanski, *Gold nanoparticle detection and quantification in therapeutic MV beams via pair production.* Phys Med Biol, 2021. **66**(6): p. 064004.
12. Allison, J., et al., *Recent developments in Geant4.* Nuclear Instruments and Methods in Physics Research Section A: Accelerators, Spectrometers, Detectors and Associated Equipment, 2016. **835**: p. 186-225.
13. Champion, C. and C. Le Loirec, *Positron follow-up in liquid water: II. Spatial and energetic study for the most important radioisotopes used in PET.* Physics in Medicine & Biology, 2007. **52**(22): p. 6605-6625.
14. Attix, F.H., *Introduction to Radiological Physics and Radiation Dosimetry*. 1986, New York: Wiley-VCH.
15. Ullom, J.N. and D.A. Bennett, *Review of superconducting transition-edge sensors for x-ray and gamma-ray spectroscopy.* Superconductor Science and Technology, 2015. **28**(8): p. 084003.
16. Marsili, F., et al., *Detecting single infrared photons with 93% system efficiency.* Nature Photonics, 2013. **7**: p. 210-214.
17. Zang, K., et al., *Silicon single-photon avalanche diodes with nano-structured light trapping.* Nat Commun, 2017. **8**(1): p. 628.
18. Llopart, X., et al., *Timepix, a 65k programmable pixel readout chip for arrival time, energy and/or photon counting measurements.* Nuclear Instruments and Methods in Physics Research Section A: Accelerators, Spectrometers, Detectors and Associated Equipment, 2007. **581**(1-2): p. 485-494.
19. Irwin, K.D. and G.C. Hilton, *Transition-edge sensors*, in *Cryogenic Particle Detection*. 2005, Springer. p. 63-149.

20. Llopart, X., et al., *Timepix4, a large area pixel detector readout chip which can be tiled on 4 sides providing sub-200 ps timestamp binning.* Journal of Instrumentation, 2022. **17**(01).
21. Wu, Z., et al., *Functionalized Hybrid Iron Oxide-Gold Nanoparticles Targeting Membrane Hsp70 Radiosensitize Triple-Negative Breast Cancer Cells by ROS-Mediated Apoptosis.* Cancers (Basel), 2023. **15**(4).
22. Zhang, Z., *Quantitative analysis of cellular uptake of nanoparticles using the Langmuir adsorption model.* Journal of Physical Chemistry B, 2014. **118**(46): p. 13196-13204.
23. Coleman, P.G., *Positron Annihilation Spectroscopy*, in *Positron Beams and Their Applications*. 2000, World Scientific. p. 151-178.
24. Dorfman, K.E., F. Schlawin, and S. Mukamel, *Entangled photon pairs and spectroscopy.* Nature Photonics, 2019. **13**(10): p. 694-701.
25. Avachat, A.V., et al., *Potential of positronium imaging for tissue characterization using PALS in total-body PET.*
26. Champion, C., *Track structure simulation of ionizing particles in water: A theoretical tool for nanodosimetry.* Physics in Medicine & Biology, 2007. **52**(23): p. 6429-6444.
27. Charlton, M. and J.W. Humberston, *Positron Physics*. 2001: Cambridge University Press.
28. Deutsch, M., *Three-Quantum Decay of Positronium.* Physical Review, 1951. **83**(4): p. 866-867.
29. Green, R.E. and A.T. Stewart, *Angular Correlation of Photons from Positron Annihilation in Light Metals.* Physical Review, 1955. **98**(2): p. 486-491.
30. Harpen, M.D., *Positronium: Review of symmetry, conserved quantities and decay for the radiological physicist.* Medical Physics, 2004. **31**(1): p. 57-61.
31. Lynn, K.G. and W.E. Frieze, *Positron annihilation spectroscopy.* Annual Review of Materials Science, 1984. **14**: p. 339-372.
32. Mogensen, O.E., *Positron Annihilation in Chemistry*. 1995: Springer.
33. Moskal, P., *Positronium imaging with the J-PET tomograph.* Nature Communications, 2021. **12**: p. 5658.
34. Moskal, P., et al., *Positronium image of the human brain in vivo.* Science Advances, 2024. **10**: p. eadp2840.
35. Moskal, P., et al., *Feasibility study of the positronium imaging with the total-body PET scanner.*
36. Moskal, P., et al., *Positronium imaging with the novel multiphoton PET scanner.* Science Advances, 2021. **7**(42): p. eabh4394.
37. Moskal, P., et al., *Positronium in medicine and biology.* Nature Reviews Physics, 2020. **2**: p. 716-728.
38. Moskal, P., E. Stępień, and S. Niedźwiecki, *Positronium and quantum entanglement imaging.* Nature Reviews Physics, 2021. **3**(8): p. 534-536.
39. Moskal, P., et al., *Positronium image of the human brain in vivo.* Science Advances, 2024. **10**: p. eadp2840.
40. Karp, J.S., et al., *PennPET Explorer: Design and preliminary performance of a whole-body imager.* Journal of Nuclear Medicine, 2020. **61**(1): p. 136-143.
41. Bordes, J., et al., *First Detailed Study of the Quantum Decoherence of Entangled Gamma Photons.* Phys Rev Lett, 2024. **133**(13): p. 132502.
42. Moodley, C. and A. Forbes, *Advances in Quantum Imaging with Machine Intelligence.* Laser & Photonics Reviews, 2024. **18**(8).

43. Schuld, M. and F. Petruccione, *Quantum machine learning*, in *Supervised Learning with Quantum Computers*. 2018, Springer. p. 1-45.